\documentclass[conference]{IEEEtran}

\usepackage{graphicx} 
\usepackage{spverbatim}
\usepackage{enumitem}
\usepackage{booktabs, tabularx} 
\usepackage[caption=false]{subfig}
\usepackage{svg}       
\usepackage{placeins}
\usepackage{multirow}
\usepackage{array}
\usepackage{tikz}
\usepackage{tcolorbox}
\usepackage{xcolor} 
\usepackage{enumitem}
\usepackage{adjustbox}
\usepackage{wrapfig}
\usepackage{helvet}
\usepackage{wasysym}
\usepackage{hyperref}
\usepackage{breakurl}
\usepackage{multicol}
\usepackage{footmisc}

\definecolor{palebluegray}{rgb}{0.86, 0.92, 0.98}
\definecolor{palegreengray}{rgb}{0.82, 0.91, 0.78}
\definecolor{palerosegray}{rgb}{0.88, 0.79, 0.83}
\definecolor{palemelon}{rgb}{0.96, 0.68, 0.}
\definecolor{brightblue}{rgb}{0.85, 0.93, 0.99}

\newcommand{\textbox}[1]{%
\begin{tcolorbox}[colback=palebluegray, colframe=palebluegray, boxrule=0.5pt, left=.5mm, right=.5mm, top=.2mm, bottom=.2mm]
    \small #1
    \vspace{-3pt}
\end{tcolorbox}
}

\newcommand{\textboxB}[1]{%
\begin{tcolorbox}[colback=palegreengray, colframe=palegreengray, boxrule=0.5pt, left=.5mm, right=.5mm, top=.2mm, bottom=.2mm]
    \small #1
    \vspace{-3pt}
\end{tcolorbox}
}

\newcommand{\textboxC}[1]{%
\begin{tcolorbox}[colback=palerosegray, colframe=palerosegray, boxrule=0.5pt, left=.5mm, right=.5mm, top=.2mm, bottom=.2mm]
    \small #1
        \vspace{-3pt}
\end{tcolorbox}
}

\definecolor{dkgreen}{HTML}{006400}

\newcommand{\smalltt}[1]{{\small\texttt{#1}}}

\newcolumntype{L}[1]{>{\raggedright\arraybackslash}p{#1}} 
\newcolumntype{C}[1]{>{\centering\arraybackslash}p{#1}} 
\newcolumntype{M}[1]{>{\centering\arraybackslash}m{#1}}

\newcommand{\boxedtext}[2][]{%
    \fcolorbox{gray!50}{white!95!gray}{%
        \parbox{0.9\linewidth}{%
            \centering
            \color{black}
            \ifx&#1&%
            \else
            \textbf{#1}\\[0.5em]
            \fi
             {\fontfamily{rm}\selectfont \textit{#2}} 
        }%
    }%
}

\title{Supporting Software Maintenance with \\ Dynamically Generated Document Hierarchies}
\author{
\IEEEauthorblockN{Katherine R. Dearstyne}
\IEEEauthorblockA{\textit{Computer Science and Engineering} \\
\textit{University of Notre Dame}\\
Notre Dame, IN, USA \\
kdearsty@nd.edu}
\and
\IEEEauthorblockN{Alberto D. Rodriguez}
\IEEEauthorblockA{\textit{Computer Science and Engineering} \\
\textit{University of Notre Dame}\\
Notre Dame, IN, USA\\
arodri39@nd.edu}
\and
\IEEEauthorblockN{Jane Cleland-Huang}
\IEEEauthorblockA{\textit{Computer Science and Engineering} \\
\textit{University of Notre Dame}\\
Notre Dame, IN, USA \\
JaneClelandHuang@nd.edu}
}

\date{October 2023}

\begin{document}
 \maketitle
\begin{abstract}
Software documentation supports a broad set of software
maintenance tasks; however, creating and maintaining high-quality, multi-level software documentation can be incredibly time-consuming and therefore many code bases suffer from a lack of adequate documentation. We address this problem through presenting HGEN, a fully automated pipeline that leverages LLMs to transform source code through a series of six stages into a well-organized hierarchy of formatted documents.  We evaluate HGEN both quantitatively and qualitatively. First, we use it to generate documentation for three diverse projects, and engage key developers in comparing the quality of the generated documentation against their own previously produced manually-crafted documentation. We then pilot HGEN in nine different industrial projects using diverse datasets provided by each project. We collect feedback from project stakeholders, and analyze it using an inductive approach to identify recurring themes. Results show that HGEN produces artifact hierarchies similar in quality to manually constructed documentation, with much higher coverage of the core concepts than the baseline approach. Stakeholder feedback highlights HGEN's commercial impact potential as a tool for accelerating code comprehension and maintenance tasks. Results and associated supplemental materials can be found at \url{https://zenodo.org/records/11403244}.
\end{abstract}

\section{Introduction}
\label{sec:intro}
Software documentation supports a broad set of software maintenance tasks such as impact analysis, change analysis, requirements validation, safety assessment, and new developer onboarding \cite{documentation-1,documentation-2,documentation-3,DBLP:conf/re/GotelF95,DBLP:conf/se/MaroSS19}, yet, creating and maintaining consistent multi-level software documentation and its associated trace links is incredibly time-consuming \cite{ DBLP:conf/icse/Cleland-HuangGHMZ14,DBLP:books/daglib/p/LuciaMOP12,rath2018traceability}. The process of documenting, defining, and maintaining documentation that describes the implemented system is often viewed as overly burdensome by developers and stakeholders. This perception leads to the documentation process being ignored, delayed, or inadequately sustained  \cite{4400153,DBLP:journals/software/MaderJZC13,DBLP:conf/icse/RempelMKC14}, especially in startups and small companies where speed is often prioritized over comprehensive requirements engineering processes \cite{paternoster_software_2014, giardino_software_2016}. Consequently, despite the many benefits of a systematic software documentation process, many code bases suffer from a lack of adequate documentation \cite{robillard-1}. 

While there have been advancements in automating certain types of software documentation, such as API specifications or the continuous deployment of embedded software documentation (\cite{openapi, doxygen}), efforts to automate the generation of comprehensive, multi-layered artifacts describing system features remain underexplored. 
With the advancements of large language models (LLMs) and their generative capabilities, there is now a path towards generating multi-layered, just-in-time software documentation; however, the challenge is in ensuring that the documentation correctly represents the underlying code base, is readable, understandable, well formatted, and properly organized so that it is useful to practitioners maintaining software systems \cite{softwaredoc-icse2020, on_demand_documentation}. 
In pursuit of this goal, we present HGEN, an automated pipeline that generates multi-layer hierarchy of documentation, comprised of artifacts such as low-level design descriptions, as well as sub-system and system-level requirements formatted according to the norms of the currently adopted life-cycle process.  HGEN not only constructs these artifacts but also generates trace links that connect them into a meaningful hierarchy, providing well organized documentation, designed to effectively support diverse software maintenance activities. We provide examples to the generated documentation for two open source datasets\footnote{\label{example-docs} Example of generated documentation: \\ \\ \textbf{Dronology}: \url{app.safa.ai/demo?version=a05d072b-163c-4ba5-a248-0683d1e2dda5&to=/project} \\ \\ \textbf{JOC}: \url{app.safa.ai/demo?version=99965515-cbc1-43e9-b834-4815f22bd2e6&to=/project}.}.

\begin{figure*}
  \centering
  \includegraphics[width=\linewidth]{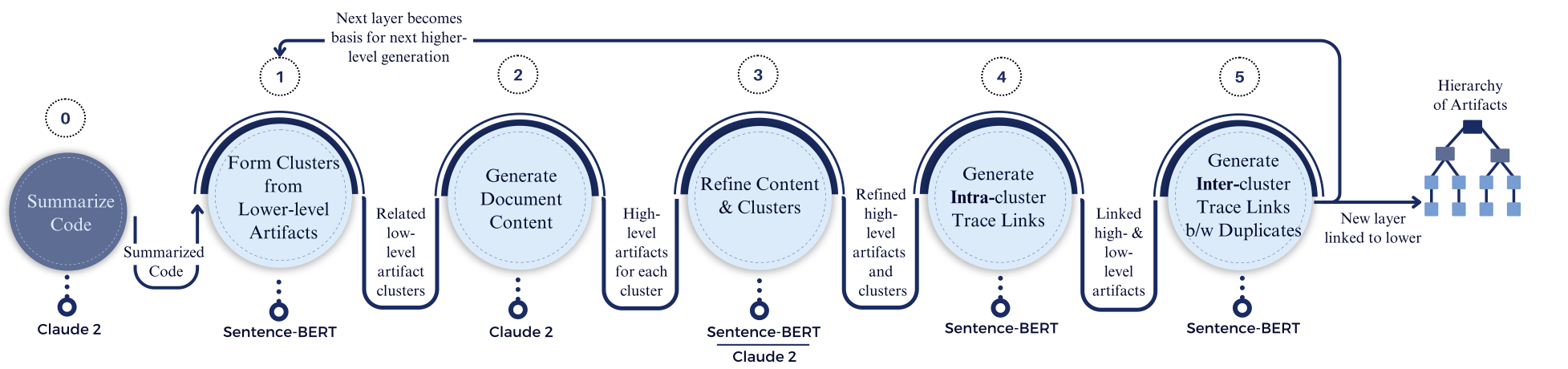}
  
  \caption{The HGEN Process utilizes a pipeline to produce each layer of the documentation hierarchy. The lowest layer accepts source code as input and generates a natural language summary (Step 0). Steps 1-5 form a pipeline, which is used to generate each subsequent layer, thereby incrementally constructing a hierarchy of progressively higher-level artifacts formatted according to the norms of the current software development process. For each step in the pipeline, we show the underlying AI models used to support the transformation of lower-level to higher-level documentation.}
  \label{fig:HGen-process}
  \vspace{-12pt}
\end{figure*}

This paper first describes the HGEN process, providing a simple running example taken from the open-source gaming domain. We then report results from two studies evaluating HGEN in which we first assessed the quality of the HGEN generated hierarchy for three different projects, and then used it to generate documentation for nine industrial pilot projects using our partners' project data. In the first study, we recruited a key developer from each of the three projects to compare HGEN's generated documentation against their own project's manual documentation and against an off-the-shelf LLM baseline. For each project, we systematically evaluated the quality of the documentation by assessing the individual quality of each generated artifact, the overall coverage of concepts, and the relationships between layers.  In the second study, we used HGEN to generate documentation for source code provided by our partners and then performed a think-aloud study in which we collected their feedback and analyzed it using an inductive approach. 

The remainder of this paper is laid out as follows.  Section \ref{sec:process} presents our process for generating the software artifact hierarchy. Section \ref{sec:evaluation} describes the design on our experiment leading Section \ref{sec:eval_quantitative} to present our quantitative evaluation of the HGEN process and Section \ref{sec:eval_qualitative} to present the qualitative feedback.  Section \ref{sec:related} presents related work on generating software documentation, code summarization, and code understanding, while Section \ref{sec:threats} discuss threats to the validity of our study. Finally, Section \ref{sec:conclusion} summarizes the overall benefits of our approach and describes future work.

\section{HGEN Process}
\label{sec:process}

The HGEN process represents a pipeline for generating hierarchies of software 
artifacts.  As depicted in Figure \ref{fig:HGen-process}, it includes five stages where Stage 1 accepts a set of {\it lower-level} artifacts as inputs, Stages 2-4 perform internal processing, and Stage 5 produces the {\it higher-level} artifacts that constitute a new layer of documentation. This new layer is then passed as inputs into Stage 1 to restart the process for creating the next layer. The stages for generating a single layer of documentation are therefore as follows:

\begin{enumerate}[label={\footnotesize Stage \arabic*.},leftmargin=*]
\item Accept a set of lower-level natural language artifacts and perform clustering on them to identify related features and/or functionality. In the special case of the lowest-level, where the inputs are source code artifacts, perform an additional pre-processing stage (Stage 0) to generate a natural language summarization of the code. This summary serves as a proxy for the source code throughout the remaining steps. Upon the conclusion of this stage, a set of clusters of lower-level artifacts is generated as output.
\item For each cluster identified in Stage 1, generate a natural language description using the targeted artifact format (e.g., user story, feature description etc). This serves as the body of the new layer of documentation.
\item Refine the content of any artifacts that contain overlapping information to improve clarity, conciseness, and ensure each artifact focuses distinctly on one specific feature or functionality.
\item Connect these refined artifacts to the lower-level input artifacts by dynamically generating trace links. 
\item Leverage the overall perspective provided by the relationships established in Stage 4 to detect and remedy redundant artifacts, and to produce the final set of output artifacts for the current layer. 
\item If a higher-layer of artifacts is desired, pass these output artifacts as inputs to the next layer. Continue this process until all targeted layers have been generated.
\end{enumerate}

The end result is a hierarchy of software artifacts, referred to as an artifact tree. Given the transformation that occurs during the generation of a single layer, we made numerous design decisions concerning the stages of the pipeline, the tasks assigned to each stage, and the algorithmic solutions for accomplishing each task. Each stage, and the overall sequence of stages,  was designed as the result of trial-and-error in which we evaluated various techniques and their combinations. We followed a robust process based on the  Design Science methodology, in which each design iteration included problem investigation, design and validation, and implementation and evaluation \cite{DBLP:books/sp/Wieringa14}. In earlier iterations, validation was performed internally by the researchers, while in later iterations it was performed by external Software Practitioners. The final outcomes of their evaluation are reported in Section \ref{sec:evaluation} of this paper.

To support our description of the HGEN process, we've chosen a small, straightforward code repository from a CS101 project as a running example. We refer to this as HERO throughout the remainder of this paper. HERO is not associated with the paper's authors and is openly accessible at https://github.com/gbaman/QUB-CSC1011-Module-Hero-Game. Due to space constraints, we do not detail the intermediate stages of the design that led to the finalized process, and focus instead on describing the end result in the following sections. Overall, our process is designed to be model-agnostic and therefore we do not present a detailed empirical comparison of results based on different LLM model types in this paper, and discuss this decision further in Section \ref{sec:threats}. We now outline each stage in our HGEN pipeline.

\begin{figure}[h]
\textbox{{\bf Hero.java:}~This code provides the framework for a user to take control of a hero character within a digital game. Upon initialization, the hero is placed in a starting location on a virtual map. Lists of crimes for the hero to address and playable characters they can select are automatically generated. The user is then able to view the hero's character details and current status. As the user navigates the hero through the game world and engages in activities to resolve crimes, their total action value increases. Periodically checking this action value triggers different game states - once a threshold is reached, the user achieves victory and the gameplay loop restarts from the main menu. The user can also adjust their hero's action directly to progress the story at will. Throughout, the code integrates the hero character with the overall game system to immerse the user in an interactive experience where they guide the actions and challenges of their virtual protagonist.}
\vspace{-3pt}
\caption{As part of our running example, HGEN summarizes the HERO source code in the preprocessing Stage 0.}
\label{fig:hero-summary}
\vspace{-12pt}
\end{figure}
 
\subsection{Stage 0: Code Summarization} 
The lowest level of the documentation hierarchy starts with source-code, and therefore a pre-processing step is applied to summarize the code into natural language. This step serves two key purposes. First, it allows us to transform source code into natural language comparable to the input artifacts of all other documentation layers.  Second, the summary has higher information density and less redundancy than raw code. This enables a larger amount of information to be conveyed within a single context window of the LLM, thereby enhancing its capacity to comprehend a broader scope of the system. Summarization tasks are best performed using generative models; therefore, we opted to use Anthropic's Claude 2.0 model, which returns similar results to OpenAI's GPT-4 \cite{gpt4} and has a large context window of 100-k tokens \cite{claude}. We prompted Claude to summarize the source code by (i) initially outlining the functionality provided to the user by the code, and (ii) then creating a polished summary that explains how the code supports the described user behavior. The resulting summarized output becomes the starting input for HGEN, representing the initial tier in the documentation hierarchy. The summary dynamically generated by HGEN for $Hero.java$ in the HERO code-base is depicted in Figure \ref{fig:hero-summary}.

\subsection{Stage 1: Form clusters from Lower-level Artifacts}
\label{sec:clustering}

We adopted a multi-technique clustering approach with the following internal steps, labeled C1-C8.

\begin{figure}[htbp]
  \centering
  \resizebox{\columnwidth}{!}{%
    \includegraphics{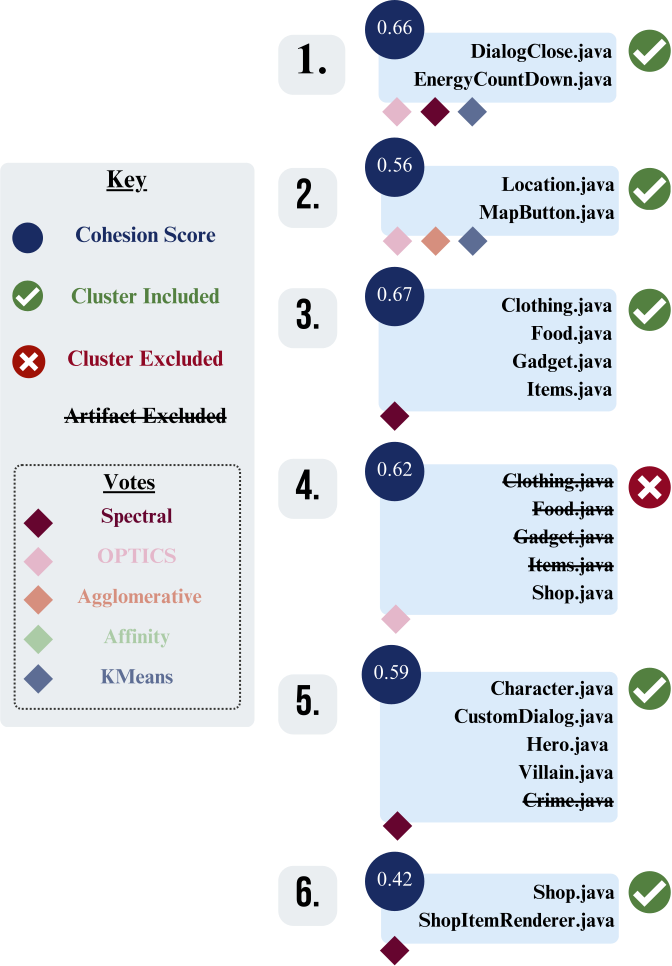} 
  }
  \caption{In Stage 1, HGEN uses multiple clustering algorithms to produce and initial set of clusters from the source code summaries, and then performs a series of filtering, ranking, and cleansing steps on the clusters.}
  \label{fig:cluster_example}
  \vspace{-18pt}
\end{figure}

\begin{enumerate}[label={\footnotesize C\arabic*.},leftmargin=*]
\item {\it Preprocessing:} 
We start by converting the natural language artifacts into embeddings using the Sentence-BERT transformer. This choice is driven by the model's capacity to encode entire sentences rather than relying on word-level encoding, as well as its consistent performance across a diverse range of tasks. As a result, Sentence-BERT is used for all transformations to embeddings throughout the remainder of this paper.
\item{\it Multi-Technique Clustering: } Early experimentation showed various unsupervised clustering algorithms each had their own unique strengths and limitations. We therefore ultimately adopted a consensus-based approach and included five different techniques to achieve diversity of cluster size, outlier detection, and geometric considerations \cite{sklearn_clustering}. These techniques were OPTICS \cite{ankerst_optics_1999}, Spectral \cite{yu_multiclass_nodate}, Agglomerative \cite{sasirekha_agglomerative_2013}, Affinity Propagation \cite{dueck_affinity_2009} and K-means \cite{arthur_k-means_2007}. In this step, each technique was used to individually cluster the vector representations of the input artifacts, producing a diverse set of candidate clusters.

\item {\it Filter by Size} The set of generated clusters were highly diverse but included overlapping and redundant clusters. We therefore applied the following filtering steps, starting by eliminating two types of clusters:

\begin{enumerate}[label={-},leftmargin=*]
\item {\it Singletons: }Temporarily set aside singleton clusters containing only one artifact. 
\item {\it Large Clusters: } Discard large clusters containing five or more artifacts, as these tend to  inhibit the LLM's ability to identify and extract finer details. While the decision to remove large clusters limits the potential for constructing higher-level abstractions across larger artifact groups, we partially address this later in Step 3, by allowing clusters to re-form.
\end{enumerate}

\item \textit{Cluster Scoring: } We assign an importance value to each remaining cluster as follows: \vspace{-4pt}
\begin{equation}
    importance = (\alpha \cdot \log(s) + h) \cdot v
    \vspace{-8pt}
\end{equation}

where:
\begin{itemize}
    \item $h$ is the cohesion score for the cluster, 
    \item $v$ is the voting score from the five clustering techniques,
    \item $s$ represents the cluster size,
    \item $\alpha$ is the weight applied to the cluster size.
\end{itemize}

The voting score ($v$) is computed by counting the number of times the exact cluster, with the same input artifacts, appears in the candidate pool.

Cohesion ($c$) is computed by averaging the cosine similarity of each artifact's embedding to all its neighbors within a cluster as follows:
    \begin{equation}
        cohesion = \frac{2}{N(N-1)} \sum_{i=1}^{N} \sum_{j=1, j\neq i}^{N} \cos(\theta_{ij})\label{eq:cohesion}
    \end{equation}

Finally, the size ($s$) metric is used to compensate for the tendency for smaller clusters to have higher cohesion. It is computed as the log of the number of input artifacts, weighted by a small constant ($alpha$). The use of logarithm moderates the impact of larger clusters, ensuring that their contribution to the importance score grows at a decreasing rate and prevents them from disproportionately dominating the score due to their size alone. 
    
\item \textit{Cluster Ranking: } Clusters are then ranked in descending order by importance score. Fig. \ref{fig:cluster_example}, depicts several clusters generated from the code base arranged by their respective importance scores.  While clusters 1 and 2 have lower cohesion scores than cluster 3, their overall score places them higher in the ranking.

\item \textit{Cluster Cleansing: } Artifact outliers that deviate by 1.5 standard deviations or more from the average similarity to their neighbors are removed from their clusters to eliminate dissimilar artifacts from the cluster. For example, Crime.java is removed from cluster 5 in Fig.\ref{fig:cluster_example}.

\item \textit{Cluster Selection: } Next, we iterate through the clusters in order of their importance to determine whether to select them for the final set. At each iteration, we consider the cluster possessing the next highest importance score (termed the $Focus Cluster$), alongside the set of clusters already chosen for inclusion (referred to as the $Inclusion Set$). Given the initial prevalence of overlapping or redundant clusters in our consensus-based approach, we first assess whether the $Focus Cluster$ contains artifacts not already present in the $Inclusion Set$. After removing any artifacts shared with clusters in the $Inclusion Set$, we admit the $Focus Cluster$ only if it maintains a size of two or more artifacts and has a cohesion score greater than or equal to the top 75\% of clusters. This process is exemplified in Cluster 4 from Fig. \ref{fig:cluster_example}. This cluster contains four artifacts already included in Cluster 3, so each of these artifacts are removed, leaving the cluster with only one unique artifact. As a result, it fails to meet the size threshold and is therefore excluded from the final cluster set.

\item \textit{Handle Orphans}
Finally, we check for artifacts not assigned to a cluster. For each orphaned artifact, we identify its most similar cluster by computing the average cosine similarity between the orphan and all members of each cluster. If the similarity is close to the cluster's overall cohesion score (within 0.1), it indicates that the orphan can be added without reducing the overall cohesion of the cluster. As a result, the orphan is incorporated into that cluster. Finally, any unplaced orphans are retained as singleton clusters. 
\end{enumerate}

\subsection{Stage 2: Generate Documentation Content}
Prior studies have shown the importance of well-formatted documentation \cite{softwaredoc-icse2020}; therefore this stage focuses on formatting the generated artifacts according to the stakeholder's needs. For example, a user might wish to generate an agile documentation hierarchy composed of source code (lowest layer), user stories (middle layer), and epics (top layer); or they might wish to generate a traditional hierarchy composed of source code, design specifications, and multiple layers of requirements. In this stage we prompt the LLM to format the output of the desired artifacts by specifying (i) the output artifact type, (ii) the desired format of the artifact,  and (iii) the targeted number of document artifacts to be generated from the current cluster. 

The artifact type is specified by the user, while the format can either be predefined by the user or generated by the LLM in a separate context window. In this study, we used the latter approach to increase the degree of automation. Given that most LLM's pre-training data contains examples of diverse common artifact types, we can simply prompt Claude to generate a standardized format for the artifact type. For instance, Claude generated the following user story template: ``As a [type of user], I want to [action or goal] so that [reason or benefit]".

Finally, we define the number of high-level artifacts (\smalltt{n\_targets}) to be generated for each cluster by considering two factors. First, \textit{cohesion} measures the extent to which an artifact focuses on a single topic. Seemingly, clusters with low cohesion typically encapsulate more topics and require more higher-level elements. We therefore compute ``concept diversity'' as the inverse of the cohesion score (cf Eq. \ref{eq:cohesion}), normalized so that the maximum ``concept diversity' for the project equals 1. Second, the amount of information within a cluster's artifacts plays a key role in determining the number of higher-level artifacts required. We estimate the \textit{information density} of the cluster by comparing the size of its artifacts to the average size of all artifacts of the same type.  Finally, we calculate the number of targeted artifacts (i.e., \smalltt{n\_targets}) as the product of ``concept diversity'' and ``information density''. To promote the emergence of a tree-like documentation structure, we impose a constraint that \smalltt{n\_targets} must be greater than 50\% and less than 100\% of the current cluster's artifact count.

Returning to the HERO example, we determine that Cluster 5 (see Figure \ref{fig:cluster_example}) requires three higher level artifacts. The four code files have a total of 730 LOC (lines of code), while the average file in this layer has 109 LOC. We estimate information density to be approximately 6.7 ($730/109$), reflecting the complex game logic contained within these core character-related classes. Normalized concept diversity is computed as $0.56$, leading to \smalltt{n\_targets} being three artifacts i.e., by computing and truncating $0.56 \times$ $6.7$.  
Figure \ref{fig:generations-example} shows the three subsequent user stories generated for Cluster 5 in our example.


\begin{figure}[h]
\textbox{{\bf [US1] Customize Character Name and Image:}~As a player, I want to be able to \underline{customize my character's name} and image so that I can personalize my gameplay experience.}
\textbox{{\bf [US2] View Character Inventory and Money:}~As a player, I want to be able to view my character's inventory and money so that I can make informed decisions when interacting with the game world.}
\textbox{{\bf [US3] Progress Character Through Story:}~As a player, I want to be able to commit crimes and take heroic actions that will progress my character through the game's story and scenarios.}
\caption{User stories generated from HERO source code during Stage 2. All 3 user stories were produced from the same cluster of source artifacts (see Cluster 5 from Figure \ref{fig:cluster_example}).}
\label{fig:generations-example}
\end{figure}

\subsection{Stage 3: Refine Content \& Clusters}
Because automatic clustering may not always match human judgment, some level of conceptual overlap across clusters is inevitable, resulting in duplicated content in the generated artifacts. Stage 3 addresses this issue by reducing duplicated content and refining the artifact clusters through three steps, labeled D1-D3.

\begin{enumerate}[label={\footnotesize D\arabic*.},leftmargin=*]
    \item \textit{Duplicate Identification}: To identify potential duplicates, we cluster the generated artifacts using the algorithm described in Section \ref{sec:clustering}. This creates groups of similar artifacts that are currently spread across different clusters. The most cohesive clusters are those most likely to contain duplicated content and thus are identified as {\it duplicate clusters}.  For example, in HERO, US1 (Fig. \ref{fig:generations-example}) is a generated artifact which is detected as similar to other character-related user stories from different clusters (US4, US5) (see Figure \ref{fig:duplicate-cluster-example}).  
    
\begin{figure}[h]
\textbox{{\bf [US1] Customize Character Name and Image}}
\textboxB{{\bf [US4] Customize Character Identity:}~As a player who wants an immersive role playing
experience, I want to be able to \underline{customize a character with a name} and choose to be a hero or villain so that I can define my virtual identity in the world}
\textboxC{{\bf [US5] Character Entity Templates for Game Testing:}~As a game developer, I want the system to allow defining
character entities via reusable templates and validated
testing so that playable characters can be reliably
generated with consistent expected behaviors for use in
game scenarios.}
\caption{Generated user stories for HERO that were clustered together in Stage 3. Although each user story originated from a different cluster in Stage 2,  they were clustered together in Stage 3 due to their shared focus on character customizations.}
\label{fig:duplicate-cluster-example}
\end{figure}

     \item \textit{Duplicate Content Identification}: 
     In this step, our aim is to determine what source artifacts led to the overlapping content so they can be re-clustered together. We identify the source artifacts contributing to the overlap by selecting those with the highest semantic similarity to the parent. Then, a new cluster is formed containing the selected source artifacts for each generated artifact in the duplicate cluster.

    \item \textit{Re-generation}: 
    At this stage, each duplicate cluster has identified the source artifacts containing the overlapping content. Now, our goal is to give the LLM a chance to regenerate new artifacts based on this focused context. Given the set of source artifacts, we repeat Stage 2 in order to generate a fresh set of artifacts centered around the core theme. These new artifacts replace those in the duplicate clusters.  In our example, the overlapping artifacts were re-generated as shown in \ref{fig:regeneration-example}, where each artifact is now focused on a more distinct topic.

\end{enumerate}

\begin{figure}[h]
  \textbox{{\bf [US1*] Character Customization:} As a player, I want to name and customize the appearance of my character so that I can roleplay a unique persona in the game world.}
  \textbox{{\bf [US4*] Play as Hero or Villian:} As a player, I want the option to play as either a hero or villain so that I can experience different perspectives when interacting with the game systems.}
  \caption{Refined User Stories for HERO during Stage 3 }
  \label{fig:regeneration-example}
\end{figure}

\subsection{Stage 4: Generate \textbf{Intra}-cluster Trace Links}

\begin{figure}[pb]
  \centering
  \resizebox{\columnwidth}{!}{%
    \includegraphics{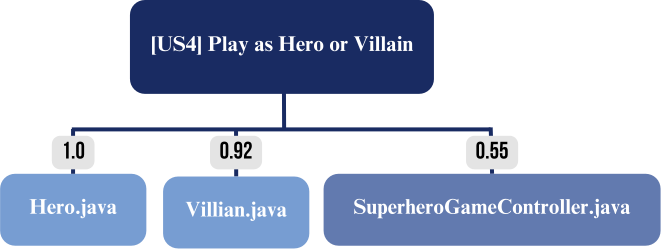} 
  }
  \caption{Example of trace link generations for US4 from the HERO example.}
  \label{fig:trace-link-example}
\end{figure}
Once the new artifacts have been generated, we need to connect them via trace links to the current layer. Given that a single cluster can produce multiple higher-level artifacts, we cannot assume  every lower-level artifact in a cluster should link to each of the resulting higher-level ones. Consequently, we create trace links only between artifacts that demonstrate strong semantic similarity using standard automated tracing techniques. First, we generate embeddings for the higher-level artifacts and use these to calculate their cosine similarity with each low-level artifact from their originating clusters. We scale each cluster's scores using min-max scaling so that the highest score is adjusted to 1. We consider the variability in scores across different clusters, and only generate links where the similarity score is within two standard deviations of the maximum normalized score. Typically, this results in a cutoff of approximately 0.8. However, if no links are generated for a lower-level artifact, we establish a link with the higher-level artifact that has the greatest similarity.

For example, the trace links established for US4 (Figure \ref{fig:regeneration-example}) are depicted in Figure \ref{fig:trace-link-example}. Following scaling, both $Hero.java$ and $Villain.java$ attain high similarity scores and are consequently linked to the user story. Although $SuperheroGameController.java$ receives a considerably lower score, its similarity to US2 exceeds its scores with all other user stories, allowing it to trace to US2 as well.

\subsection{Step 5: Generate \textbf{Inter}-cluster Trace Links between Duplicates}
\label{sec:duplicates}

\begin{figure}[b]
  \centering
  \resizebox{\columnwidth}{!}{%
    \includegraphics{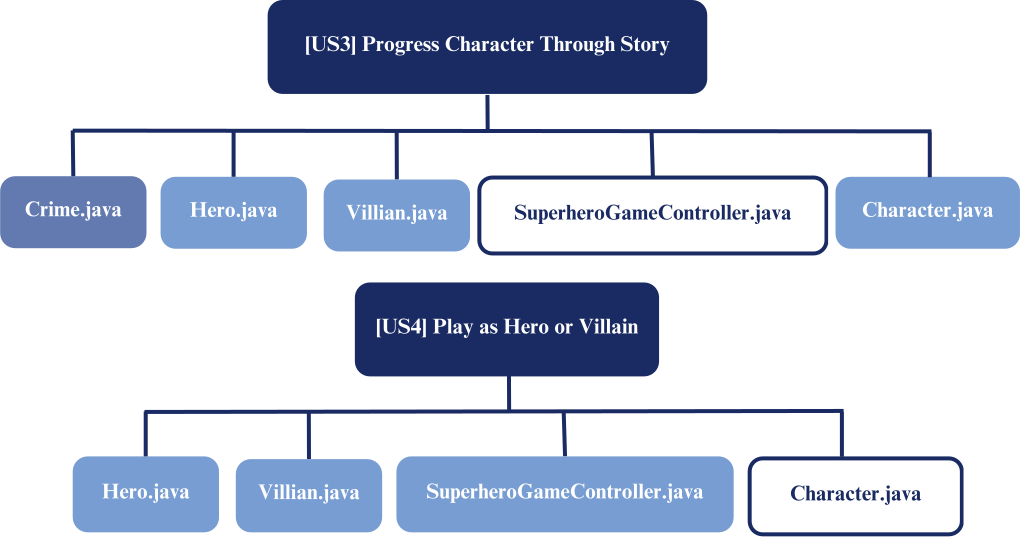} 
  }
  \caption{Trace links selected for US3 and US4 after they were flagged containing overlapping content in Stage 5.  Both user stories gain 1 additional trace link from the other (shown in white). }
  \label{fig:duplicate-detection-example}
\end{figure}
In this step, we identify any remaining artifacts with overlapping content, which also aids in detecting potential trace links between clusters. First, we compute the cosine similarity between each pair of generated artifacts, marking pairs with similarity scores more than two standard deviations above the mean as potential duplicates. For each duplicate pair, designated as A and B, we consider whether any of B's trace links should also trace to A and vice verse. Trace links are formed if the similarity score between B and the child is of a similar strength between A and the child (i.e., within a difference of 0.1). 

If two pairs of highly similar artifacts result in trace links with identical child artifacts, it signifies that the pair are likely duplicates. In such cases, we remove one of the duplicates, as the risk of losing crucial information is significantly reduced. 



An illustration of this re-tracing process can be seen in Figure \ref{fig:duplicate-detection-example}. Artifacts that were originally traced to each user story are shown in blue. After re-tracing, each user story gains an additional trace link, represented in white. Notably, US3 possesses one trace link ($Crime.java$) not linked to US4; however, in this example, both US3 and US4 are retained after this stage, as US3 is linked to $Crime.java$, while US4 is not.

\section{Experiment Design}
\label{sec:evaluation}

Evaluating the effectiveness of documentation hierarchies is complex because there is no single ground-truth solution \cite{techdoc1,techdoc2,techdoc3}. While it is tempting to use automated assessment techniques such as BLEU, METEOR, ROUGE, CIDEr, and SPICE, to detect overlapping terms across documents, Hu et al., showed that the metrics do not align with human judgment about the quality of documentation  \cite{Hu_Chen_Wang_Xia_Lo_Zimmermann_2022}. Therefore, our evaluation primarily leveraged human judgment. In our first study, we recruited a knowledgeable project stakeholder to systematically compare HGEN's performance against their own project documentation in three different projects. While in the second study, we conducted nine industry pilot studies using HGEN and elicited general feedback on its performance. In this section we describe the first study.

\subsection{Projects}
\label{sec:projects}
\begin{table}
\centering
\caption{The first study evaluated the quality of manually constructed documentation, HGEN, and a baseline approach. Types and numbers of artifacts are depicted for each project. Open-Source datasets are annotated with an asterisk (*).}
\label{tab:projects}
\begin{tabular}{@{}p{4.5cm}p{3.5cm}@{}}
\toprule
\textbf{Dronology *} \cite{DBLP:journals/corr/abs-1804-02423} \newline 
Dronology is an Open-Source small Unmanned Aerial System (sUAS) written in Java. It provides a platform for controlling and coordinating multiple sUAS to support search-and-rescue, surveillance, and scientific data collection missions.
&
\vspace{-12pt}
\raisebox{-\height}{\raggedright\includegraphics[width=3.5cm]{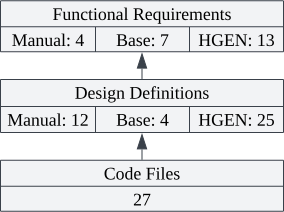}}  \\ 

\vspace{2pt}\textbf{SAFA} \cite{DBLP:conf/kbse/RodriguezNDC22,DBLP:journals/software/Cleland-HuangAV21}\newline 
SAFA is a software documentation management platform that leverages live traceability to build a knowledge graph and support change impact analysis. Our industry collaborators provided access to closed-source client-side source code.
&
\raisebox{-\height}{\centering\includegraphics[width=3.5cm]{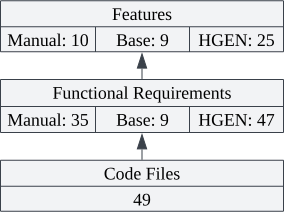}}\\

\vspace{2pt}\textbf{Jack of Clubs *}  \newline 
Jack of Clubs is a re-creation of Ace of Spades, a voxel-based first-person shooter game. The creator gave us access to the manually created user stories and epics, and we are releasing them to the public as part of this paper.
&
\raisebox{-\height}{\centering\includegraphics[width=3.5cm]{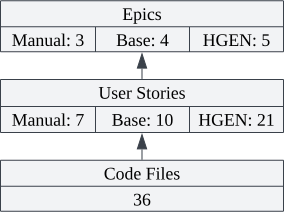}}\\ 
\bottomrule
\end{tabular}
\vspace{-12pt}
\end{table}

The three projects are summarized in Table \ref{tab:projects}. Our three projects represent diverse domains and cover both traditional and agile development processes. Each project included source code and at least two types of natural language artifacts, organized into layers, and connected by trace links. The Dronology is a UAV project that includes 423 Java files, 211 Design Definitions, and 99 Requirements respectively. SAFA is a software safety tool and includes  242 Vue source-code files, 262 functional requirements, and 101 features. Finally, Jack of Clubs is a first-person shooter game and includes 36 C++ files, 7 user stories, and 3 epics. Each project had an available Senior Developer with prior experience in developing documentation, to serve as the project expert.

\subsection{Techniques under Comparison}
Our study involved three different treatments including (i) the HGEN generated documentation, (ii) a baseline LLM approach, and (iii) the project documentation previously developed manually by project personnel. The HGEN documentation was generated for the three projects following the steps outlined in Section \ref{sec:process} of this paper, while the manually constructed project documentation was provided by the original project stakeholders as referenced in Table \ref{tab:projects}. 

We also created a baseline LLM approach for comparison purposes. We started with the identical set of summarized code as the HGEN processs and used the same LLM (Claude 2.0) to generate comprehensive documentation for each artifact type used in HGEN. As with HGEN, we repeat the process for each layer, and connect artifacts across layers by generating embeddings (using Sentence-BERT) for both the lower and higher-level artifacts. Trace links are established between artifact  across the two layers if the normalized cosine similarity score is greater than 0.7. The major difference between the BASELINE and HGEN is the use of clustering techniques in HGEN as well as the later refinement steps.

Due to the size of Dronology and SAFA, we extracted a subset of source-code files and their generated documentation so that the project experts could conduct a more in-depth analysis of each artifact. In each case, the project expert identified the most critical source files, and the study included those files and their linked documentation.

\section{Quantitative Comparison of the Documentation Quality}
\label{sec:eval_quantitative}

To evaluate the quality of the generated documentation we addressed the following research questions. 

\begin{table}[]

  \addtolength{\tabcolsep}{-3pt}
    \centering
    \small

\caption{Evaluation guidelines for assessing documentation quality.}
\label{tab:quality_metrics}
\small
\begin{tabular}{|L{1.2cm}|L{1.6cm}|L{5.3cm}|}
\hline
\textbf{Group} & \textbf{Metric} & \textbf{Metric Description} \\ \hline
Language & Readability & How easily a user can understand the information provided. \\ \cline{2-3}

& Appropriate-ness & Language appropriateness with respect to the artifact's technical level. \\
\hline
Content & Conciseness & How brief, yet clear the system is described.\\
\cline{2-3}
& Importance & Considers the importance of the information provided.
\\
\hline
Effective-ness & Usefulness & Considers how useful this documentation is to the project expert. \\
\cline{2-3}
& Helpfulness & Consider how helpful the documentation is for understanding its children. \\
\hline
\end{tabular}
\vspace{-12pt}
\end{table}


\begin{itemize}[leftmargin=*]

\item \textbf{RQ1. Artifact Quality: How does the quality of individual machine-generated artifacts compare to that of expert-made artifacts?}
We addressed this RQ by asking our project experts to evaluate the language, content, and effectiveness of each artifact as proposed by Hu et al., \cite{Hu_Chen_Wang_Xia_Lo_Zimmermann_2022}. {\it Language} included readability and appropriateness, {\it content} included conciseness and importance, and {\it effectiveness} included usefulness and helpfulness. The definitions provided to our human assessors are summarized in Table \label{tab:quality_metrics} and provided in our supplemental material.
    
\item \textbf{RQ2. Coverage: To what extent are concepts that appear in the original documentation {\it covered by} the generated documentation? }
To answer this RQ we measured concept coverage \cite{lawson2008use,SEBoK_concept} and evaluated the extent to which concepts appearing in the original documentation appeared in the generated documentation. Examples of concepts in HERO might be the ability to select different characters or purchase specific items at the shop.
    
\item \textbf{RQ3. Relationships: How effectively does the machine-generated documentation build appropriate parent-child relationships inherent in multi-layer data structures? } Relationships are represented by trace links, and we therefore evaluated them using standard traceability metrics of recall and precision for the generated artifact tree \cite{DBLP:conf/icse/Cleland-HuangGHMZ14}
\end{itemize}

\subsection{RQ1. Artifact Quality} 
\label{sec:rq1}
Project experts comparatively evaluated artifacts in their own project for (a) the manually constructed documentation, (b) HGEN, and (c) the baseline approach using a qualitative rubric associated with each quality in Table \ref{tab:quality_metrics}. Evaluations were performed against a Likert scale ranging from 1-5, where 1 signified low quality and 5 represented high quality. They were allowed to move freely between the different types of artifacts during this process. 

\noindent{\it Analysis of results:}~Due to non-normal score distributions, we employed the Mann–Whitney U test, which is non-parametric, to test if there was a notable difference in score distributions between each of the three treatments. Further, to account for multiple comparisons in our study—18 tests across six metrics and three documentation groups (Human-made, \emph{Baseline}, \emph{HGEN})—we use the Holm-Šidák method to adjust our p-values to control for error rates, ensuring a reliable statistical analysis when comparing documentation quality across groups.
Table \ref{tab:hypothesis_tests} reports the mean scores for the six quality attributes assigned by the project experts across all three projects, as well as the adjusted p-values. It highlights instances where the null hypothesis is rejected ($p < 0.05$), suggesting that scores from one distribution tend to have higher scores than the other. 

\begin{table}
\centering
\caption{Results of Mann–Whitney U tests. Cases in which the null hypothesis is rejected (i.e., where ($p < 0.05$)) are highlighted and depict cases where one treatment outperformed the other with respect to the quality attribute.}
\label{tab:hypothesis_tests}

\begin{tabular}{p{1cm}ccc}
\hline
\multicolumn{4}{c}{\emph{Human vs. Baseline}} \\
\hline
\textbf{Metric} & \textbf{Human Mean} & \textbf{Baseline Mean} & \textbf{Corrected P-value} \\
\hline
Readability & 3.90 & \colorbox{brightblue}{\bf 4.44} & \textbf{0.002445} \\
Appropriateness & 3.70 & 4.00 & 0.214476 \\
Conciseness & 4.30 & 4.44 & 0.596927 \\
Importance & 4.10 & 4.35 & 0.191984 \\
Usefulness & 3.51 & 3.67 & 0.794294 \\
Helpfulness & 3.82 & 3.88 & 0.990378 \\
\end{tabular}

\begin{tabular}{p{1cm}ccc}
\hline
\multicolumn{4}{c}{\emph{Human vs. HGEN}} \\
\hline
\textbf{Metric} & \textbf{Human Mean} & \textbf{HGEN Mean} & \textbf{Corrected P-value} \\
\hline
Readability & 3.90 & \colorbox{brightblue}{4.38} & \textbf{0.000104} \\
Appropriateness & 3.70 & \colorbox{brightblue}{4.16} & \textbf{0.001493} \\
Conciseness & 4.30 & 4.32 & 0.618787 \\
Importance & 4.10 & 4.33 & 0.095525 \\
Usefulness & 3.51 & \colorbox{brightblue}{4.14} & \textbf{7.09e-06} \\
Helpfulness & 3.82 & \colorbox{brightblue}{4.21} & \textbf{0.002445} \\
\end{tabular}

\begin{tabular}{p{1cm}ccc}
\hline
\multicolumn{4}{c}{\emph{Baseline vs. HGEN}} \\
\hline
\textbf{Metric} & \textbf{Baseline Mean} & \textbf{HGEN Mean} & \textbf{Corrected P-value} \\
\hline
Readability & 4.44 & 4.38 & 0.990378 \\
Appropriateness & 4.00 & 4.16 & 0.462604 \\
Conciseness & 4.44 & 4.32 & 0.990378 \\
Importance & 4.35 & 4.33 & 0.990378 \\
Usefulness & 3.67 & \colorbox{brightblue}{4.14} & \textbf{0.020646} \\
Helpfulness & 3.88 & 4.21 & 0.172422 \\
\hline
\end{tabular}
\vspace{-12pt}
\end{table}

\noindent{\it Discussion of results:~} The comparison between the human constructed documentation and the baseline approach returned comparable scores on all metrics except \emph{Readability}, which was higher for the baseline approach.  The same comparison between human  and HGEN documentation showed that HGEN returned higher quality scores across four of the six metrics: \emph{Readability}, \emph{Appropriateness}, \emph{Usefulness}, and \emph{Helpfulness}. On the other hand in a direct comparison of HGEN versus the Baseline method, the only significant difference observed was for  \emph{Usefulness}, with HGEN's higher score indicating that project experts thought its documentation was more useful than the baselines. These results confirm the findings of previous studies that LLMs are able to produce software documentation of comparable quality to humans \cite{Bencheikh_Höglund_2023}. Notably, the main difference between HGEN and baseline is in the way HGEN constructs the hierarchy and not in the way it generates individual documents.

\subsection{RQ2: Coverage} We asked each expert to evaluate concept coverage by identifying concepts in the manual documentation and checking whether they were adequately reflected in the generated documentation. We then computed the proportion of concepts from the original documentation that were addressed in each of the generated documentations.


\begin{table}[t]
  \centering
  \caption{Percentage of Concepts from Original Documentation Captured Per HGEN Version, as determined by the Project Experts (E1-E3).}
  \label{tab:coverage}
    \addtolength{\tabcolsep}{-3.5pt}
  \begin{tabular}{c|c|cc|cc}
  \hline
  \multirow{2}{*}{\textbf{Project}} &\multirow{2}{*}{\textbf{ID}}& \multicolumn{2}{c|}{\textbf{Baseline}} & \multicolumn{2}{c}{\textbf{HGEN}} \\ \cline{3-6} 
   && \textbf{\% Covered} & \textbf{Covered by} & \textbf{\% Covered} & \textbf{Covered by} \\ \hline \hline
  Dronology & E1& 6.3\% & 9.1\% & 87.5\% & 28.9\% \\ \hline
  SAFA & E2 & 37.8\% & 38.9\% & 84.4\% & 43.1\% \\ \hline
  JOC & E3& 50.0\% & 35.7\% & 100\% & 38.5\% \\ \hline
  \end{tabular}
  \vspace{-12pt}
\end{table}

\noindent{\it Analysis of results:}~Results are reported in Table  \ref{tab:coverage} in the column labeled  (\% Covered). We also report the percentage of artifacts in the generated documentation that included these concepts (Covered by). The ``Covered by" percentages for both the Baseline and HGEN documentation suggest that many artifacts focus on concepts that were not highlighted in the manual version.

\noindent{\it Discussion of results:} HGEN demonstrates a notable increase in concept coverage compared to the baseline approach across all three projects, capturing twice as many concepts in both JOC and SAFA, and an impressive 80\% increase in the case of Dronology. Additionally, a larger portion of the HGEN-generated documentation, as indicated by the ``Covered by" metric, centers on core concepts from the manual documentation, particularly in the case of Dronology. Given that the experts did not identify duplicate artifacts, it appears that both the Baseline and HGEN uncovered project aspects not emphasized in the manual documentation. Matched with the increase in `helpfulness' returned by experiments for RQ1, it appears that the additional information could be helpful to project stakeholders. 

\subsection{RQ3: Relationships} To evaluate the quality of relationships within each generated hierarchy, we developed a basic tracing tool which visualized the generated documentation tree and allowed project experts to approve or decline existing links, adding new links if needed. Each expert performed this task twice -{}- once for HGEN and once for the Baseline approach, resulting in their version of a ground truth solution for each generative technique.  We then evaluated the generated trees for HGEN and Baseline against the modified ground truth version for each one and computed  mean average precision (mAP), precision, and recall for each project using standard formulas \cite{DBLP:conf/icse/Cleland-HuangGHMZ14}. To compute mAP, which assesses the extent to which correct links appear at the top of a ranked list, we ordered the links according to their original cosine similarity scores. In addition, we assessed the number of orphan artifacts generated by each approach, as this aspect had emerged as a key distinguishing factor through discussions with the three experts. Table \ref{tab:tracing} presents these results.

\begin{table}
\centering
\label{tab:tracing}
\caption{Traceability Accuracy Metrics for Generative Approaches}
\begin{tabular}{llcccc}
\hline
\textbf{Project} & \textbf{Approach} & \textbf{mAP} & \textbf{Precision} & \textbf{Recall} & \textbf{\# Orphans} \\ \hline \hline
\multirow{2}{*}{Dronology} & Baseline & 84.5\% & 47.2\% & 89.5\% & 17 \\
 & HGEN & 94.0\% & 56.3\% & 93.4\% & 0 \\ \hline
\multirow{2}{*}{SAFA} & Baseline & 91.9\% & 49.2\% & 100\% & 28 \\
 & HGEN & 94.5\% & 54.3\% & 98.4\% & 9 \\ \hline
\multirow{2}{*}{JOC} & Baseline & 95.5\% & 67.3\% & 74.5\% & 11 \\
 & HGEN & 96.7\% & 81.4\% & 80.2\% & 1 \\ \hline
\end{tabular}
\end{table}

\noindent{\it Discussion of results:~}The relatively high mAP scores (80\% to 95.5\%) and recall scores  (47.2\% - 81.4\%) indicate that Sentence-BERT was able to capture a range of semantic similarities between artifacts. This is likely attributable to the LLM's use of lower-level artifacts for generating higher-level ones, thereby creating a shared vocabulary.  However, we also observed a significantly higher number of orphans in the Baseline approach versus HGEN, which could have lowered recall whilst increasing precision. HGEN's lower orphan count suggests that it's enhanced clustering techniques enabled it to capture the concepts in the low-level artifacts at higher levels of abstraction, identifying concepts that might otherwise have been overlooked.

\subsection{Qualitative Feedback}
We also asked each expert a number of open-ended questions including having them describe the most and least valuable characteristics of the documentation. A full list of these questions can be found in the paper's supplemental section. We briefly summarize their feedback.

The experts acknowledged the quality of the baseline's individual artifacts, which, in the case of E3, was identified as more readable than the project expert's own documentation. However, both E1 and E2 identified that the baseline version lacked comprehensiveness, clarity, and accurate prioritization of information compared to the manual documentation. Furthermore, its sparse generations resulted in the creation of ``redundant" parents highlighting the \emph{Baseline's} tendency to establish 1-1 relationships between its initial and final layer generations.

All experts preferred the HGEN version over the baseline method. E2 and E3 stated that HGEN tended to produce more detailed and comprehensive information that provided ``helpful" relationships and dependencies, with links generally ``making sense." 

Despite their preference for HGEN, the experts noted some shortcomings. E2 mentioned that HGEN missed some obvious links between clusters and did not reflect the same organizational structure for conceptualizing code as they had used. This difference in structure lead both E1 and E2 to favor their own documentation over HGEN's, although E2 acknowledged HGEN's was ``more useful for teaching someone about my system," suggesting that preference depended on the documentation's intended use. Meanwhile, E3 appreciated HGEN's detailed information and preferred it over their own.

\section{Industrial Pilot Studies}
\label{sec:eval_qualitative}
We now present the results of our industrial pilot studies which focused purely on the HGEN solution. 

\subsection{Study Method} The nine pilots were conducted in seven different companies on eight unique projects as depicted in Table \ref{tab:pilots}. Seven of them used data provided by the company, and two (marked with an asterisk) used open-source project data \cite{Autowarefoundation}.  For each project we applied HGEN to the source code to generate documentation.

To assess the effectiveness of the documentation, we conducted interviews with nine industrial partners. During these interviews, partners were prompted to share their candid thoughts on the documentation's quality, usefulness, and suggestions for potential improvements. After obtaining permission, we recorded the sessions and utilized an AI tool to automatically transcribe the recordings. Two of the paper's authors independently extracted all relevant quotes, used an inductive approach to encode each quote, and then worked together to discuss the code and to sort them into categories. We did not assess inter-rater agreement as this activity was performed collaboratively.

\subsection{Analysis of Feedback}
Our analysis identified three clear themes and six sub-categories associated with documentation quality, prospective use cases, and recommendations for improvements.

\subsubsection{Documentation Quality}
Two sub-themes emerged for documentation quality. The first addressed {\it information accuracy and coverage} and focused on whether the generated documentation conveyed crucial information about the source code without errors. Feedback was generally positive. Participant P9 expressed strong satisfaction, stating, ``everything I can read corresponds exactly to the reality," while P1 remarked, ``I feel like they are a good representation of what we are doing." However, one participant, P2, noted a discrepancy where an external tool was mentioned despite not being implemented in the code. On the other hand they pointed out that the tool was referenced in the code as a prerequisite, which likely misled the LLM.  P4 stated that they couldn't find any errors at all, and confirmed that it captured all essential information, stating that ``I couldn't identify any aspect missing from it."

The second theme focused on {\it clarity and structure}, including  readability and understandably of the documentation. P4 observed that the documentation was well written, and was ``probably better than I could do." P8 specifically praised the hierarchical organization of the documentation, expressing appreciation for the fact that it provided ``a summary at every level of depth...[and] every level of extraction."

\subsubsection{Prospective Use cases}
Participants also focused on how HGEN could benefit their respective companies, with three specific use cases emerging, all of which are highly pertinent to software maintenance.

Five participants (P2, P6, P7, P8, P11) highlighted HGEN's potential for {\it comprehending complex systems} lacking sufficient documentation, especially in scenarios where the original authors are unavailable. Of these, P7 underscored its use when ``nobody knows anything about [the code]'' stating that ``you run it through your system, and then it's a lot more potentially clear, and people can understand what was going on." P11 felt that it would be particularly advantageous for ``tackling some pretty legacy stuff." Participants P6 and P7 enthusiastic about HGEN's time-saving capabilities, with P6 stating that without documentation, understanding a system might take a month, whereas with the HGEN documentation ``I have an idea maybe within days. So it's definitely a big help on that one." 

Four participants (P1, P2, P3, P6) highlighted the value of HGEN for expediting the onboarding of new developers. P2 stated that they would no longer need ``to spend tons of time giving the engineer an overview of the code base", and P3 echoed this sentiment, stating that it was ``a whole lot better than me setting an intern down with some piece of code that I got and telling him. 'Hey do your best buddy. We'll talk to you in four months'".

Finally, two participants (P7, P11) discussed the use of HGEN for regulatory compliance. P7 observed that generating documentation would simplify the process, stating, ``instead of going to the code and trying to figure out what to provide to them'' (i.e., regulators), ``[HGEN] would be somewhat easier." P11 said that HGEN would have assisted them in a previous government project, in which they were required to maintain an ``enormous'' and comprehensive list of requirements, which was challenging to maintain. They believed that utilizing HGEN might have made the task more manageable.

\subsubsection{Potential Improvements}
Two participants (P1, P3) recommended new features aimed at enhancing HGEN's utility. P1 noted that the generated documentation failed to include details from external libraries, suggesting that relevant parts of these libraries would provide context for the LLM. P3 proposed adapting HGEN for instances where high-level documentation already exists, suggesting that generated documentation could bridge the gap between this high-level overview and the code. Finally, P1 was intrigued by the potential for extending HGEN to provide incremental support for documentation alongside code development.

\begin{table}
    \centering
     \caption{HGEN was used in nine industrial Pilot Projects from multiple domains with the source code layer written in diverse programming languages.}
  
    \label{tab:pilots}
      \addtolength{\tabcolsep}{-3pt}
    \begin{tabular}{|c|l|c|c|c|c|l|c|}
    \hline
    \multirow{2}{*}{\textbf{ID}} &
    \multirow{2}{*}{\textbf{Category}} &
    \multicolumn{3}{c|}{\textbf{Use Case}} &
    \multirow{2}{*}{\textbf{ID}} &
    \multirow{2}{*}{\textbf{Lang.}} & \multirow{2}{*}{\textbf{Input Files}}\\ \cline{3-5} 
    && RE. & LG & OB & && \\ \hline
      
    \multirow{2}{*}{C1} & \multirow{3}{*}{Enterprise} & \CIRCLE &\CIRCLE & &P1 & C\# & 1,049  \\ \cline{3-8}
    && \CIRCLE &\CIRCLE && P2 & C++ & 181 \\ \hline
    C2&Automotive*& & && P3 & C++ \textasteriskcentered  & 242 \textasteriskcentered \\    
    \hline
    
    C3 & SAAS & & \CIRCLE && P4 & TS / JS& 264 \\  \hline
    C4 & IT Services & \CIRCLE&&& P5 & Java& 197\\  \hline
    C5 & Aerospace & &\CIRCLE && P6 & C & 345  \\  \hline
    C6 & Education & &&\CIRCLE& P7 & Java & 643\\  \hline
    C7 & Automotive* & &&& P8 & \textasteriskcentered & \textasteriskcentered \\ \hline
    C8 & IT Services & &\CIRCLE&& P9 & Go / TS & 65 \\  \hline
    \multicolumn{8}{c}{*Pilot conducted using Open-Source Systems}\\  \multicolumn{8}{c}{RE=Reverse Engineering, LG=Legacy Documentation, OB=Onboarding}\\
    \multicolumn{8}{c}{JS= Java Script, TS = Type Script}
    \end{tabular}
    \vspace{-12pt}
   
\end{table}

\subsection{Discussion of Results}
The feedback provided by project stakeholders highlighted several benefits and potential applications of HGEN as well as some places for improvement. Most importantly, it demonstrated that HGEN was capable of generating high quality documentation hierarchies for an extremely wide range of software projects. However, these results are based on a pilot study, and therefore feedback is based on stakeholders perspectives of HGEN's utility rather than on its adoption in practice. Nevertheless, this is an important first step in validating HGEN for deployment on industrial projects.

\section{Related Work}
\label{sec:related}
Our work is informed by the groundwork laid by prior research in the areas of automated document generation \cite{on_demand_documentation}. We discuss most closely related work in three specific areas.

First, there is a large body of work in automating code-level documentation and API specifications for various frameworks \cite{doxygen, openapi, sphinx, swaggerui, javadocs, django, springboot, fastapi, nodejs}. Our research focuses on generating hierarchical, multi-level documentation for higher-level system abstraction. We build on the emerging results showing that LLMs can generate a variety of software requirements and documentation. Dvivedi et al. showed that LLM-based models often surpass human documentation in inline, function, and file level code documentation \cite{dvivedi2023comparative}. Likewise, Bencheikh and Höglund's demonstrated that LLMs can generate software requirements \cite{Bencheikh_Höglund_2023}, while Xie et al. use it to generate code specifications \cite{Xie_Yoo_Jiang_Kim_Tan_Zhang_Lee_2023}. Our work aims to enhance the existing capabilities of LLMs to create diverse documentation types at multiple abstraction levels. In the private sector, companies like swimm.io have investigated documentation automation, but claim that full automation is infeasible due to the difficulty of integrating business logic, design decisions, and other external elements \cite{rosenbaum_will_2023}. While this is important, our work demonstrates the benefits of automation as a component of the documentation process.

Research in source code understanding has predominantly focused on generating detailed, low-level explanations. Notable works include Srihara et al.'s file-level code summarizations using natural language generators \cite{10.1145/1858996.1859006}, and methods by Robillard, Burden, Moreno, among others, for creating concise method and class summaries \cite{robillard-1, robillard-2, Burden, moreno}. Others have explored parameter-level comments \cite{5970165} and context-enhanced summaries \cite{7181703, McBurney_McMillan_2014}, including method functionality, purpose, and usage \cite{7181703}. HGEN, in contrast, targets higher-level, language-agnostic documentation, leveraging LLMs to summarize code across most major programming languages.

The pursuit of ubiquitous software traceability serves as a foundational inspiration, drawing from extensive prior research dedicated to the development and refinement of automatic trace link generation across various software domains \cite{systems_and_software_traceability, grand_challenges, DBLP:books/daglib/p/LuciaMOP12, rath2018traceability, DBLP:conf/icse/RahimiC19}. Despite substantial progress in software traceability \cite{bert_pl, bert_nl}, challenges in achieving high accuracy remain, especially in data-scarce areas \cite{DBLP:conf/icse/Cleland-HuangGHMZ14, traceability_nn}. Recently, LLMs like GPT3, GPT4, and Claude have demonstrated potential in improving trace link accuracy in such scenarios \cite{rodriguez2023understanding, gpt4, claude, prompts}. Our approach integrates advances from BERT-based models for trace link generation and LLMs for documentation generation, with a novel clustering strategy to increase trace accuracy.

\section{Threats to Validity}
\label{sec:threats}
Our work includes several important threats to validity. With respect to construct validity, our first study focused on three projects only. However, we partially mitigated the threat to generalizability by selecting  projects from diverse domains, which encompassed both open-source and closed-source code for different sized project. We then applied HGEN to industrial project data, and the feedback from project experts indicated that it was effective across all domains. However, our pilot studies were based on feedback elicited from an interactive demonstration, and while this provides valuable insights, an additional study is needed of its use over time in industrial projects. 
In a threat to external validity, the HGEN pipeline is quite complex, developed following much trial and error and its replication is complex. To mitigate this, we ensured repeatability in the hierarchy generation process by minimizing randomness and maintaining a closely-deterministic approach in our pipeline. Multiple runs were conducted to confirm consistent results. Further, we have prepared a fully functioning system that is accessible via a web application.\footref{example-docs}  We also provide all study materials in the supplemental materials. 


\section{Conclusion}
\label{sec:conclusion}

This paper proposes HGEN, an LLM-based approach to automatically generate hierarchies of requirements documentation from source code. HGEN builds upon the recent successes of LLMs by engineering a process that addresses some of the deficiencies identified with the unaided models. Our evaluation, designed to target three critical aspects of the documentation, supports existing literature that LLM-generated documentation can match or exceed the quality of that written by humans. We also show that HGEN is able to capture meaningful relationships across varied artifact levels and can identify nearly all of the core concepts found in expert-produced documentation, showing a considerable enhancement over the baseline LLM. These results indicate that HGEN could substantially reduce the time and effort needed for comprehensive documentation creation, thereby aiding in software maintenance tasks. Moreover, HGEN presents potential for automating additional aspects of requirements engineering, paving the way forward towards ubiquitous documentation and traceability.

\bibliographystyle{IEEEtran}
\bibliography{software.bib}

\end{document}